# Dissipative delay range analysis of coupled differential-difference delay systems with distributed delays

Qian Feng, Sing Kiong Nguang

*Department of Electrical and Computer Engineering, The University of Auckland, Auckland 1010, New Zealand*

**Abstract**

This paper proposes methods to handle the problem of delay range stability analysis for a linear coupled differential-difference system (CDDS) with distributed delays subject to dissipative constraints. The model of linear CDDS contains many models of linear delay systems as special cases. A novel Liapunov-Krasovskii functional with non-constant matrix parameters, which are related to the delay value polynomially, is applied to derive stability conditions. By constructing this new functional, sufficient conditions in terms of robust linear matrix inequalities (LMIs) can be derived, which guarantee range stability of a linear CDDS subject to dissipative constraints. To solve the resulting robust LMIs numerically, we apply the technique of sum of squares programming together with matrix relaxations without introducing any potential conservatism to the original robust LMIs. Furthermore, the proposed methods can be extended to solve delay margin estimation problems for a linear CDDS subject to prescribed dissipative constraints. Finally, numerical examples are presented to demonstrate the effectiveness of the proposed methodologies.

*Keywords:* Coupled Differential-Difference Systems, Range stability, Dissipativity, Sum of Square programming

## 1. Introduction

Functional differential equations Hale & Lunel (1993) are able to characterize a dynamical process whose behavior is affected by its past values, i.e a dynamical system conditioned by delay effects. To analyze the stability property of such system, however, is non-trivial due to its infinite dimensional nature. Two major directions, which are based on either time Gu *et al.* (2003) or frequency domain Michiels & Niculescu (2014), have been investigated to provide solutions to characterize how delays affect the stability of a system.

For a linear delay system, the information of its stability can be obtained by analyzing its corresponding spectrum. Many different approaches Gu *et al.* (2003); Kharitonov *et al.* (2009) have been developed in frequency domain, which can provide almost a complete stability characterization when the delay systems exhibit certain structures. For more complex delay structures such as distributed delays with general kernels, the numerical schemes in Breda *et al.* (2005, 2014); Vyhlídal & Zítek (2014) can produce reliable results verifying system stability with given point-wise delay values, which suffer almost no conservatism. Furthermore, the method in Gumussoy & Michiels (2011) allows one to calculate the value of $\mathcal{H}^\infty$ norm of a delay system with known point-wise delay values. However, to the best of our knowledge, none of the existing spectral based approaches can handle the problem of delay range stability analysis subject to performance objectives Scherer *et al.* (1997) for linear delay systems. Namely, to test whether a delay system is stable and simultaneously dissipative Briat (2014) for all delay value $r \in [r_1, r_2]$ with respect to a supply function, where the exact delay value $r$ is unknown but bounded by $r_1 \leq r \leq r_2$ with known values $r_2 > r_1 > 0$.

On the other hand, constructing Krasovskii functional Gu *et al.* (2003); Briat (2014) has been applied as a standard approach in time domain to analyze the stability of a delay system. Many different functionals

*Email addresses:* `qfen204@aucklanduni.ac.nz` (Qian Feng), `sk.nguang@auckland.ac.nz` (Sing Kiong Nguang)

(see Gu *et al.* (2003); Fridman (2014); Briat (2014) and the references therein) have been proposed among existing literature Gu *et al.* (2001); Feng & Nguang (2016b) to analyze the problem of point-wise delay stability. Compared to its frequency domain counterparts, time domain approaches may be more adaptable to handle range stability analysis with performance objectives, though only sufficient conditions can be derived. In Gouaisbaut *et al.* (2013); Ariba *et al.* (2017), the results concerning the range stability of a linear discrete delay system are presented based on the principle of quadratic separation. On the other hand, the solutions of the same problem have been proposed based on constructing Krasovskii functionals in Seuret & Gouaisbaut (2015); Gyurkovics & Takács (2016). However, no results, based on the Krasovskii approach, concerning range stability analysis have been proposed when distributed delays are considered.[1] On the other hand, almost all existing Krasovskii functionals in literature are based on constant matrix parameters, which is a very conservative choice when it comes to range stability analysis. This motives one to propose new functionals to specifically tackle the problem of range stability analysis considering performance objectives or even further potential constraints.

In this paper, we propose methodologies which allow one to conduct range stability analysis for a linear coupled differential-difference system (CDDS) Gu & Liu (2009); Li (2012); Hongfei (2015) subject to dissipative constraints. The linear CDDS model considered in this paper contains distributed delay terms with polynomials kernels, which is able to incorporate many models of time delay systems as special cases. A novel Liapunov-Krasovskii functional, with delay-dependent matrix parameters, is applied to be constructed together with a quadratic supply function to derive stability conditions. The resulting sufficient conditions, expressed in robust LMIs, are able to ensure the range stability and dissipativity of the linear CDDS over a known delay interval. To solve the robust LMIs numerically, we apply SoS programming Chesi (2010); Blekherman *et al.* (2013) with the relaxation technique in Scherer & Hol (2006) to equivalently transfer the original polynomials optimization problem into semidefinite programs with finite dimensions, without introducing any potential conservatism. Furthermore, the proposed scenario is extended to handle the problem of estimating the margin of a stable delay interval with given dissipative constraints. Finally, we also prove that the resulting stability conditions in this paper exhibit a hierarchical feasibility enhancement similar to the one in Seuret & Gouaisbaut (2015).

The paper is organized as follows. In section 2 we formulate the linear CDDS model to be analyzed in this paper. Subsequently, theoretical preliminaries are presented in section 3 which provide necessary tools to derive the main results in the following section. In section 4, the main results on range stability analysis incorporating dissipative constraints are presented, including remarks and detailed explanations. Finally, we present several numerical examples in section 5 to demonstrate the advantage of our proposed schemes.

**Notation**

The notations in this paper follow standard rules. In addition, we introduce certain new symbols for the sake of having efficient presentations. We define $\mathbb{T} := \{x \in \mathbb{R} : x \geq 0\}$ and $\mathbb{S}^n := \{X \in \mathbb{R}^{n \times n} : X = X^\top\}$. We frequently apply the notations of universal quantifier $\forall$ and the existential quantifier $\exists$ throughout the paper. $\mathcal{X}^\mathcal{Y}$ standards for the set containing all possible functions defined from $\mathcal{Y}$ onto $\mathcal{X}$. The notations $\|\mathbf{x}\|_q = \left(\sum_{i=1}^n |x_i|^q\right)^{\frac{1}{q}}$ and $\|f(\cdot)\|_p = \left(\int_\mathbb{R} |f(t)|^p \mathrm{d}t\right)^{\frac{1}{p}}$ and $\|\boldsymbol{f}(\cdot)\|_p = \left(\int_\mathbb{R} \|\boldsymbol{f}(t)\|_2^p \mathrm{d}t\right)^{\frac{1}{p}}$ are the norms/seminorms associated with $\mathbb{R}^n$ and Lebesgue integrable functions spaces $\mathbb{L}^p(\mathbb{R}\,;\mathbb{R})$ and $\mathbb{L}^p(\mathbb{R}\,;\mathbb{R}^n)$, respectively. $\mathbf{Sy}(X) := X + X^\top$ is the sum of a matrix with its transpose. A column vector containing a sequence of objects is defined as $\mathbf{Col}_{i=1}^n x_i := \left[\mathbf{Row}_{i=1}^n x_i^\top\right]^\top = \left[x_1^\top \cdots x_i^\top \cdots x_n^\top\right]^\top$. The symbol $*$ is applied to denote $[*]YX = X^\top YX$ or $X^\top Y[*] = X^\top YX$. We use $\mathsf{O}_{n \times n}$ to indicate a $n \times n$ zero matrix with the abbreviation $\mathsf{O}_n$, whereas $\mathbf{0}_n$ denotes a $n \times 1$ column vector. The notations $X \succ Y$ ($X \prec Y$) is equivalent to $X - Y \succ 0$ ($X - Y \prec 0$) which means that $X - Y$ is positive (negative) definite, whereas $<$ and $>$ indicate point-wise orders. (The corresponding partial order relations of the aforementioned relations follow the same rules). The diagonal sum of two matrices and $n$ matrices are defined as $X \oplus Y = \mathsf{Diag}(X,Y) = \begin{bmatrix} X & \mathsf{O} \\ \mathsf{O} & Y \end{bmatrix}$ and $\bigoplus_{i=1}^n X_i = \mathsf{Diag}_{i=1}^n(X_i)$, respectively. Furthermore, $\otimes$ stands for the Kronecker product. Moreover, we assume the order of matrix operations as *matrix (scalars) multiplications* $> \otimes > \oplus >$ *matrix (scalar) additions*. Finally, the notion of empty matrix, which follows the same definition in Matlab (see https://au.mathworks.com/help/matlab/ref/zeros.html?requestedDomain=www.mathworks.com), is applied in this article to render our results more adaptable to the handling of different problems. All the matrix operations concerning empty matrices follow the same rules in Matlab.

---

[1] The methods proposed in Gouaisbaut & Ariba (2011) can handle polynomials distributed delay kernels. However, the approaches in Gouaisbaut & Ariba (2011) are derived not based on Krasovskii functionals, but the principle of robust control (Quadratic Separation).

## 2. Problem Formulation

In this paper the linear CDDS

$$\begin{aligned}
\dot{\boldsymbol{x}}(t) &= A_1 \boldsymbol{x}(t) + A_2 \boldsymbol{y}(t-r) + \int_{-r}^{0} A_3(r) L_d(\tau) \boldsymbol{y}(t+\tau) \mathsf{d}\tau + D_1 \boldsymbol{w}(t) \\
\boldsymbol{y}(t) &= A_4 \boldsymbol{x}(t) + A_5 \boldsymbol{y}(t-r) \\
\boldsymbol{z}(t) &= C_1 \boldsymbol{x}(t) + C_2 \boldsymbol{y}(t-r) + \int_{-r}^{0} C_3(r) L_d(\tau) \boldsymbol{y}(t+\tau) \mathsf{d}\tau + D_2 \boldsymbol{w}(t) \\
\mathsf{Col}\left(\boldsymbol{x}(0), \boldsymbol{y}(0+\cdot)\right) &= \mathsf{Col}\left(\boldsymbol{\xi}, \boldsymbol{\phi}(\cdot)\right) \in \mathbb{R}^n \times \widehat{\mathbb{C}}([-r,0]\,\mathring{,}\,\mathbb{R}^\nu)
\end{aligned} \qquad (1)$$

is considered, where $\boldsymbol{x}(t) \in \mathbb{R}^n$ and $\boldsymbol{y}(t) \in \mathbb{R}^\nu$ are the solution of the coupled differential-difference equations in (1), $\boldsymbol{w}(\cdot) \in \mathbb{L}^2(\mathbb{T}\,\mathring{,}\,\mathbb{R}^q)$ represents disturbance, $\boldsymbol{z}(t) \in \mathbb{R}^m$ is the regulated output. Furthermore, $\boldsymbol{\xi} \in \mathbb{R}^n$ and $\boldsymbol{\phi}(\cdot) \in \widehat{\mathbb{C}}([-r,0]\,\mathring{,}\,\mathbb{R}^n)$ are the initial conditions where $\widehat{\mathbb{C}}(\mathcal{X}\,\mathring{,}\,\mathbb{R}^n)$ stands for the Banach space of bounded right piecewise continuous functions with an uniform norm $\|\boldsymbol{f}(\cdot)\|_\infty := \sup_{\tau \in \mathcal{X}} \|\boldsymbol{f}(\tau)\|_2$. The dimensions of the state space matrices in (1) are determined by the indexes $n; \nu \in \mathbb{N}$ and $m; q \in \mathbb{N}_0 := \mathbb{N} \cup \{0\}$. Moreover, $L_d(\tau) := \boldsymbol{\ell}_d(\tau) \otimes I_\nu$ where $\boldsymbol{\ell}_d(\tau) \in \mathbb{R}^{d+1}$ contains polynomials at each row up to degree $d \in \mathbb{N}_0$. $A_3(r) \in \mathbb{R}^{n \times \varrho}$ and $C_3(r) \in \mathbb{R}^{m \times \varrho}$ are polynomials matrices of $r$ with $\varrho = (d+1)\nu$. $r$ is a constant but with unknown and bounded values as $r \in [r_1, r_2]$, where the values of $r_2 > r_1 > 0$ are known. Finally, it is assumed $\rho(A_5) < 1$ which ensures the input to state stability of $\boldsymbol{y}(t) = A_4 \boldsymbol{x}(t) + A_5 \boldsymbol{y}(t-r)$ Gu & Liu (2009) where $\rho(A_5)$ stands for the spectra radius of $A_5$. Since $\rho(A_5) < 1$ is independent from $r$, thus this condition ensures the input to state stability of $\boldsymbol{y}(t) = A_4 \boldsymbol{x}(t) + A_5 \boldsymbol{y}(t-r)$ for all $r > 0$.

**Remark 1.** Many delay related systems can be modeled by (1). See Gu & Liu (2009); Feng & Nguang (2016b) and the references therein. In comparison with the CDDS model in Gu & Liu (2009), (1) takes disturbances into account and incorporates distributed delay terms with polynomials kernels at both the state and output. In terms of real-time applications, the structures of $A_3(r)$, $C_3(r)$ can be justified by the fact that the distributed delay gain matrices can be related to the numerical values of $r$ Gu *et al.* (2016). See a representative example by the Example 2 in Gu *et al.* (2001).

## 3. Preliminaries

*3.1. Legendre polynomials*

Without losing generalities, we assume in this paper that $\boldsymbol{\ell}_d(\tau) = \mathsf{Col}_{i=0}^d \ell_i(r,\tau)$ in (1) consists of Legendre polynomials Seuret & Gouaisbaut (2014, 2015); Seuret *et al.* (2015); Feng & Nguang (2016a)

$$\ell_d(r,\tau) := \sum_{k=0}^d \binom{d}{k}\binom{d+k}{k}\left(\frac{\tau}{r}\right)^k = \sum_{k=0}^d \binom{d}{k}\binom{d+k}{k}\tau^k r^{-k}, \ \forall d \in \mathbb{N}_0, \ \forall \tau \in [-r,0], \qquad (2)$$

with $\int_{-r}^0 \boldsymbol{\ell}_d(\tau) \boldsymbol{\ell}_d^\top(\tau) \mathsf{d}\tau = \bigoplus_{k=0}^d \frac{r}{(2k+1)}$. Note that the form of (2) is derived from the structure of Jacobi polynomials Gautschi (2004) with $\alpha = \beta = 0$ in Feng & Nguang (2016a).

Some properties of Legendre polynomials are summarized as follows.

**Property 1.** *Given $d \in \mathbb{N}_0$ and $\boldsymbol{m}_d(\tau) := \mathbf{Col}_{i=0}^d \tau^i$, then the following three properties hold for all $r > 0$.*

- $\exists! \mathsf{L}_d(\cdot) \in \left(\mathbb{R}_{[d+1]}^{(d+1)\times(d+1)}\right)^{\mathbb{R}_+}, \exists! \Lambda_d \in \mathbb{R}_{[d+1]}^{(d+1)\times(d+1)} : \forall \tau \in \mathbb{R}, \ \boldsymbol{\ell}_d(\tau) = \mathsf{L}_d(r)\boldsymbol{m}_d(\tau) = \Lambda_d \left(\bigoplus_{i=0}^d r^i\right)^{-1} \boldsymbol{m}_d(\tau)$ (3)

- $\mathsf{L}_d^{-1}(r) = \left(\bigoplus_{i=0}^d r^i\right) \Lambda_d^{-1}$ (4)

- $\exists! \acute{\mathsf{L}}_d \in \mathbb{R}^{(d+1)\times(d+1)}, \forall \tau \in \mathbb{R}, \ \dfrac{\mathrm{d}\boldsymbol{\ell}_d(\tau)}{\mathrm{d}\tau} = r^{-1}\acute{\mathsf{L}}_d \boldsymbol{\ell}_d(\tau)$ (5)

where $\mathbb{R}_{[n]}^{n\times n} := \{X \in \mathbb{R}^{n\times n} : \mathrm{rank}(X) = n\}$. and $\exists!$ *stands for the symbol of unique existential quantification.*

*Proof.* Since Legendre polynomials are linear independent, thus (3) can be easily derived based on the form of (2). By (3) and $r > 0$, (4) can be obtained. Finally, by (3), we have $\frac{\mathrm{d}\boldsymbol{\ell}_d(\tau)}{\mathrm{d}\tau} = \Lambda_d \left(\bigoplus_{i=0}^d r^i\right)^{-1} \frac{\mathrm{d}\boldsymbol{m}_d(\tau)}{\mathrm{d}\tau}$. Now it is obvious that, $\frac{\mathrm{d}\boldsymbol{m}_d(\tau)}{\mathrm{d}\tau} = \begin{bmatrix} \mathbf{0}_d^\top & 0 \\ \bigoplus_{i=1}^d i & \mathbf{0}_d \end{bmatrix} \boldsymbol{m}_d(\tau)$ for all $d \in \mathbb{N}_0$ if we define $\mathbf{0}_0$ and $\bigoplus_{i=1}^0 i$ to be $0 \times 1$ and $0 \times 0$ empty matrices, respectively. Using this relation with (3) and (4) we can obtain that

$$\frac{\mathrm{d}\boldsymbol{\ell}_d(\tau)}{\mathrm{d}\tau} = \Lambda_d \left(\bigoplus_{i=0}^d r^i\right)^{-1} \frac{\mathrm{d}\boldsymbol{m}_d(\tau)}{\mathrm{d}\tau} = \Lambda_d \left(\bigoplus_{i=0}^d r^i\right)^{-1} \begin{bmatrix} \mathbf{0}_d^\top & 0 \\ \bigoplus_{i=1}^d i & \mathbf{0}_d \end{bmatrix} \boldsymbol{m}_d(\tau)$$

$$= \Lambda_d \left(\bigoplus_{i=0}^d r^i\right)^{-1} \begin{bmatrix} \mathbf{0}_d^\top & 0 \\ \bigoplus_{i=1}^d i & \mathbf{0}_d \end{bmatrix} \left(\bigoplus_{i=0}^d r^i\right) \Lambda_d^{-1} \boldsymbol{\ell}_d(\tau) \quad (6)$$

Note that based on the final term in (6), (6) can be rewritten into

$$\frac{\mathrm{d}\boldsymbol{\ell}_d(\tau)}{\mathrm{d}\tau} = \Lambda_d \begin{bmatrix} 1 & \mathbf{0}_d^\top \\ \mathbf{0}_d & \bigoplus_{i=1}^d r^{-i} \end{bmatrix} \begin{bmatrix} \mathbf{0}_d^\top & 0 \\ \bigoplus_{i=1}^d i & \mathbf{0}_d \end{bmatrix} \begin{bmatrix} \bigoplus_{i=0}^{d-1} r^i & \mathbf{0}_d \\ \mathbf{0}_d^\top & r^d \end{bmatrix} \Lambda_d^{-1} \boldsymbol{\ell}_d(\tau)$$

$$= \Lambda_d \begin{bmatrix} \mathbf{0}_d^\top & 0 \\ r^{-1}\left(\bigoplus_{i=1}^d i\right) & \mathbf{0}_d \end{bmatrix} \Lambda_d^{-1} \boldsymbol{\ell}_d(\tau) \quad (7)$$

which proves (5). □

**Remark 2.** Consider distributed delay terms with standard polynomials kernels such as $\mathbb{R}^{n\times\nu} \ni \widehat{A}(\tau) = AM_d(\tau) = A(\boldsymbol{m}_d(\tau) \otimes I_\nu)$ and $\mathbb{R}^{m\times\nu} \ni \widehat{C}(\tau) = CM_d(\tau) = C(\boldsymbol{m}_d(\tau) \otimes I_\nu)$, where $\boldsymbol{m}_d(\tau) := \mathbf{Col}_{i=0}^d \tau^i$ and the matrices $A \in \mathbb{R}^{n\times\varrho}$, $C \in \mathbb{R}^{m\times\varrho}$ can be easily determined by the structure of $M_d(\tau)$. By the definition of $\boldsymbol{\ell}_d(\tau)$ in (2) with (3) and (4), we have $AM_d(\tau) = A\left(\mathsf{L}_d^{-1}(r) \otimes I_\nu\right) L_d(\tau)$ and $CM_d(\tau) = C\left(\mathsf{L}_d^{-1}(r) \otimes I_\nu\right) L_d(\tau)$, where $A\left(\mathsf{L}_d^{-1}(r) \otimes I_\nu\right)$ and $C\left(\mathsf{L}_d^{-1}(r) \otimes I_\nu\right)$ are polynomials matrices with respect to $r$ in line with $A_3(r)$ and $C_3(r)$ in (1). This demonstrates that the choice of Legendre polynomials $\boldsymbol{\ell}_d(\tau)$ in (2) together with the forms of $A_3(r)$ and $C_3(r)$ in (1) can handle standard polynomials matrix distributed delay terms.

*3.2. Criteria of range delay stability and dissipativity*

The following range stability criteria of CDDS can be obtained by modifying the Theorem 3 of Gu & Liu (2009).

**Lemma 1.** *Given $r_2 > r_1 > 0$ and a differential-difference system*

$$\dot{\boldsymbol{x}}(t) = \boldsymbol{f}(\boldsymbol{x}(t), \boldsymbol{y}(t+\cdot)), \quad \boldsymbol{y}(t) = \boldsymbol{g}(\boldsymbol{x}(t), \boldsymbol{y}(t+\cdot)), \quad \boldsymbol{f}(\boldsymbol{0}_n, \boldsymbol{0}_\nu(\cdot)) = \boldsymbol{0}_n, \quad \boldsymbol{g}(\boldsymbol{0}_n, \boldsymbol{0}_\nu(\cdot)) = \boldsymbol{0}_\nu(\cdot) \quad (8)$$

where $\boldsymbol{y}(t + \cdot) \in \widehat{\mathbb{C}}([-r,0]\,;\mathbb{R}^\nu)$ and $\boldsymbol{y}(t) = \boldsymbol{g}(\boldsymbol{x}(t), \boldsymbol{y}(t+\cdot))$ *is uniformly input to state stable. Then* (8) *is globally uniformly asymptotically stable at its origin for all* $r \in [r_1, r_2]$, *if there exist* $\epsilon_1; \epsilon_2; \epsilon_3 > 0$ *and a differentiable functional* $v : \mathbb{R}_+ \times \mathbb{R}^n \times \widehat{\mathbb{C}}([-r,0]\,;\mathbb{R}^\nu) \to \mathbb{T}$ *such that* $v(r, \boldsymbol{0}_n, \boldsymbol{0}_\nu(\cdot)) = 0$ *and*

$$\epsilon_1 \|\boldsymbol{\xi}\|_2^2 \leq v(r, \boldsymbol{\xi}, \boldsymbol{\phi}(\cdot)) \leq \epsilon_2 \left(\|\boldsymbol{\xi}\|_2 \vee \|\boldsymbol{\phi}(\cdot)\|_\infty\right)^2 \tag{9}$$

$$\dot{v}(r, \boldsymbol{\xi}, \boldsymbol{\phi}(\cdot)) \leq -\epsilon_3 \|\boldsymbol{\xi}\|_2^2 \tag{10}$$

*hold for all* $r \in [r_1, r_2]$ *and for all* $\boldsymbol{\xi} \in \mathbb{R}^n$ *and for all* $\boldsymbol{\phi}(\cdot) \in \widehat{\mathbb{C}}([-r,0]\,;\mathbb{R}^\nu)$, *where* $a \vee b := \max(a,b)$ *and*

$$\dot{v}(r, \boldsymbol{\xi}, \boldsymbol{\phi}(\cdot)) := \left.\frac{\mathsf{d}^+}{\mathsf{d}t} v(r, \boldsymbol{x}(t), \boldsymbol{y}(t+\cdot))\right|_{t=\tau, \boldsymbol{x}(\tau)=\boldsymbol{\xi}, \boldsymbol{y}(\tau+\cdot)=\boldsymbol{\phi}(\cdot)}, \quad \frac{\mathsf{d}^+}{\mathsf{d}x} f(x) = \limsup_{\eta \downarrow 0} \frac{f(x+\eta) - f(x)}{\eta} \tag{11}$$

*with* $\dot{\boldsymbol{x}}(t)$ *and* $\boldsymbol{y}(t+\cdot)$ *satisfying* (8).

*Proof.* The Theorem 3 of Gu & Liu (2009) is for a given $r > 0$ where $r$ is a variable of the system equation. However it can be easily extended pointwisely by treating $r$ in the system as an uncertain parameter belonging to an interval $[r_1, r_2]$ with $r_2 > r_1 > 0$. Moreover, the functions $V(\cdot), u(\cdot), v(\cdot)$ and $w(\cdot)$ in the Theorem 3 of Gu & Liu (2009) should be parameterized by $r$ in this case. Thus a corresponding range stability criteria can be obtained which can verify the stability of (8) for all $r \in [r_1, r_2]$. Following the aforementioned steps and letting the functions $u(r, \cdot), v(r, \cdot), w(r, \cdot)$ to be the quadratic functions $\epsilon_i x^2$, $i = 1, 2, 3$, Lemma 1 can be obtained accordingly. $\square$

**Definition 1** (Dissipativity). Given $r_2 > r_1 > 0$, a CDDS system

$$\dot{\boldsymbol{x}}(t) = \boldsymbol{f}(\boldsymbol{x}(t), \boldsymbol{y}(t+\cdot), \boldsymbol{w}(t)), \quad \boldsymbol{y}(t) = \boldsymbol{g}(\boldsymbol{x}(t), \boldsymbol{y}(t+\cdot)),$$

with $\boldsymbol{z}(t) = \boldsymbol{h}(\boldsymbol{x}(t), \boldsymbol{y}(t+\cdot), \boldsymbol{w}(t))$ is dissipative with respect to a supply rate function $s(\boldsymbol{z}(t), \boldsymbol{w}(t))$ for all $r \in [r_1, r_2]$, if there exists a differentiable functional $v : \mathbb{R}_+ \times \mathbb{R}^n \times \widehat{\mathbb{C}}([-r,0]\,;\mathbb{R}^\nu) \to \mathbb{R}$ such that

$$\forall r \in [r_1, r_2], \ \forall t \in \mathbb{T}: \ \dot{v}(r, \boldsymbol{x}(t), \boldsymbol{y}(t+\cdot)) - s(\boldsymbol{z}(t), \boldsymbol{w}(t)) \leq 0, \tag{12}$$

where (12) is a delay range version of the original definition of dissipativity, given $v(r, \boldsymbol{x}(t), \boldsymbol{y}(t+\cdot))$ is differentiable Briat (2014).

To incorporate dissipative constraints into the analysis of (1), a quadratic supply function

$$s(\boldsymbol{z}(t), \boldsymbol{w}(t)) = \begin{bmatrix} \boldsymbol{z}(t) \\ \boldsymbol{w}(t) \end{bmatrix}^\top \mathbf{J} \begin{bmatrix} \boldsymbol{z}(t) \\ \boldsymbol{w}(t) \end{bmatrix} \quad \text{with} \quad \mathbf{J} = \begin{bmatrix} J_1 & J_2 \\ * & J_3 \end{bmatrix} \in \mathbb{S}^{(m+q)}, \ J_1 \preceq 0 \tag{13}$$

is applied in this paper which is taken from Scherer *et al.* (1997). For specific optimization objectives included by (13) such as $\mathbb{L}^2$ gain performance ($J_1 = -\gamma^{-1} I_m$, $J_2 = \mathsf{O}_{m \times q}$, $J_3 = \gamma I_q$), see the details in Scherer *et al.* (1997).

*3.3. Some useful results on Kronecker products and integral inequality*

**Lemma 2.** $\forall X \in \mathbb{R}^{n \times m}, \ \forall Y \in \mathbb{R}^{m \times p}, \ \forall Z \in \mathbb{R}^{q \times r}$,

$$(X \otimes I_q)(Y \otimes Z) = (XY) \otimes (I_q Z) = (XY) \otimes Z = (XY) \otimes (Z I_r) = (X \otimes Z)(Y \otimes I_r). \tag{14}$$

*Proof.* (14) is the particular case of the property of Kronecker product $(A \otimes B)(C \otimes D) = (AC) \otimes (BD)$. $\square$

The following inequality has been first derived in Seuret & Gouaisbaut (2014, 2015) with different notations.

**Lemma 3.** *Given $U(r) \in \mathbb{S}^n_{\succeq 0}$ for all $r > 0$, then the inequality*

$$\int_{-r}^{0} \boldsymbol{x}^\top(\tau) U \boldsymbol{x}(\tau) \mathsf{d}\tau \geq [*] \left(r^{-1} \mathsf{D}_d \otimes U(r)\right) \left[\int_{-r}^{0} \left(\boldsymbol{\ell}_d(\tau) \otimes I_n\right) \boldsymbol{x}(\tau) \mathsf{d}\tau\right] \tag{15}$$

*holds for all $\boldsymbol{x}(\cdot) \in \mathbb{L}^2([-r,0]; \mathbb{R}^n)$ and for all $r > 0$, where $\boldsymbol{\ell}_d(\tau)$ has been defined in* (2) *and* $\mathsf{D}_d := \bigoplus_{i=0}^{d} 2i+1$.

*Proof.* Given $U(r) \in \mathbb{S}^n_{\succeq 0}$ for all $r > 0$. Let $\mathcal{K} = [-r, 0]$ and $\boldsymbol{f}(\tau) = \boldsymbol{\ell}_d(\tau)$ in Lemma 5 in Feng & Nguang (2016b), then it gives the form of the inequality (15) with a known $r$ since $\int_{-r}^{0} \boldsymbol{\ell}_d(\tau) \boldsymbol{\ell}_d^\top(\tau) \mathsf{d}\tau = r\mathsf{D}_d^{-1}$. Note that the result is naturally valid for all $r > 0$ which gives this lemma. □

**Remark 3.** Note that since $\boldsymbol{x}(\cdot) \in \mathbb{L}^2([-r,0]; \mathbb{R}^n)$ in (15) with the fact that all functions in $\boldsymbol{\ell}_d(\tau)$ are bounded, therefore all the integrals in (15) are well defined.

*3.4. Polynomials optimization*

Now we present in this subsection the foundation of polynomials optimization, which is instrumental in solving the stability conditions derived in this paper. See Scherer & Hol (2006); Chesi (2010); Blekherman *et al.* (2013) for detailed discussions on this subject.

In the following definition, we define the space of univariate polynomials matrices. For the expression of multivariate polynomials matrices, see Chesi (2010).

**Definition 2.** The space containing polynomials matrices between $\mathbb{R}^n$ to $\mathbb{R}^{p \times q}$ is defined as

$$\mathbb{R}^{p \times q}[\mathbb{R}] := \left\{ F(\cdot) \in (\mathbb{R}^{p \times q})^{\mathbb{R}} \;\middle|\; \begin{array}{c} F(x) = \sum_{i=1}^{p} Q_i x^p \quad \& \quad p \in \mathbb{N}_0 \\ \& \quad Q_i \in \mathbb{R}^{p \times q} \end{array} \right\}. \tag{16}$$

Furthermore, the degree of a polynomial matrix is defined as

$$\deg\left(\sum_{i=1}^{p} Q_i x^p\right) = \max_{i=1\cdots p} \left(\left[\mathbb{1}_{\mathbb{R}^{p \times q} \setminus \{\mathsf{O}_{p \times q}\}}(Q_i)\right] i\right) \tag{17}$$

where $\mathbb{1}_{\mathcal{X}}(\cdot)$ is the standard indicator function. This also allows us to define

$$\mathbb{R}^{p \times q}[\mathbb{R}]_d := \left\{ F(\cdot) \in \mathbb{R}^{p \times q}[\mathbb{R}] : \deg(F(\cdot)) = d \right\}, \text{ with } d \in \mathbb{N}_0 \tag{18}$$

which contains polynomials with degree $d$.

The following definition gives the space of univariate sum of square polynomials matrix. For the definition of the structure of multivariate sum of square polynomials matrix, see Scherer & Hol (2006) for details.

**Definition 3.** A polynomial in $\mathbb{S}^m[\mathbb{R}]$ is classified as a sum of square polynomial if and only if it belongs to the space

$$\boldsymbol{\Sigma}\left(\mathbb{R}; \mathbb{S}^m_{\succeq 0}\right) := \left\{ F(\cdot) \in \mathbb{S}^m[\mathbb{R}] \;\middle|\; \begin{array}{c} F(x) = \Phi(x)^\top \Phi(x) \\ \exists \Phi(\cdot) \in \mathbb{R}^{p \times m}[\mathbb{R}] \quad \& \quad p \in \mathbb{N} \end{array} \right\}. \tag{19}$$

We also define $\boldsymbol{\Sigma}_d\left(\mathbb{R}; \mathbb{S}^m_{\succeq 0}\right) := \left\{ F(\cdot) \in \boldsymbol{\Sigma}\left(\mathbb{R}; \mathbb{S}^m_{\succeq 0}\right) : \deg(F(\cdot)) = 2d \right\}$ with $d \in \mathbb{N}_0$. Finally, it is obvious to see that $\boldsymbol{\Sigma}_0\left(\mathbb{R}; \mathbb{S}^m_{\succeq 0}\right) = \mathbb{S}^m_{\succeq 0}$.

The following lemma allows one to solve SoS constraints numerically via semidefinite programming. Unlike the original Lemma 1 in Scherer & Hol (2006), we only need to consider the univariate case.

**Lemma 4.** $P(\cdot) \in \boldsymbol{\Sigma}\left(\mathbb{R}; \mathbb{S}^m_{\succeq 0}\right)$ *if and only if there exists* $Q \in \mathbb{S}^{(d+1)m}_{\succeq 0}$ *such that*

$$\forall x \in \mathbb{R}, \quad P(x) = (\boldsymbol{m}(x) \otimes I_m)^\top Q (\boldsymbol{m}(x) \otimes I_m), \tag{20}$$

*where* $\boldsymbol{m}(x) := \mathbf{Col}_{i=0}^{d} x^i$ *with* $d \in \mathbb{N}_0$.

*Proof.* Let $u(\cdot) = \boldsymbol{m}(x) := \textsf{Col}_{i=0}^{d} x^i$ in the Lemma 1 of Scherer & Hol (2006), then Lemma 4 is obtained. $\square$

**Remark 4.** When it comes to real-time calculation, one can only obtain a numerical result $Q \succ 0$ instead of $Q \succeq 0$. Consequently, the membership certificate produced by numerical calculations in reality is $P(\cdot) \in \boldsymbol{\Sigma}\left(\mathbb{R}\,\mathring{,}\,\mathbb{S}_{\succ 0}^{m}\right) \subset \boldsymbol{\Sigma}\left(\mathbb{R}\,\mathring{,}\,\mathbb{S}_{\succeq 0}^{m}\right)$.

## 4. Main Results of Dissipative Stability Analysis

In this section, the main results of dissipative stability analysis are presented.

### 4.1. A Krasovskii functional with delay dependent parameters

To analyze the stability of the origin of (1) for all $r \in [r_1, r_2]$, we consider a parameterized functional of the form

$$v(r, \boldsymbol{x}(t), \boldsymbol{y}(t+\cdot)) := \begin{bmatrix} \boldsymbol{x}(t) \\ \int_{-r}^{0} L_d(\tau)\boldsymbol{y}(t+\tau)\mathsf{d}\tau \end{bmatrix}^{\top} P(r) \begin{bmatrix} \boldsymbol{x}(t) \\ \int_{-r}^{0} L_d(\tau)\boldsymbol{y}(t+\tau)\mathsf{d}\tau \end{bmatrix}$$
$$+ \int_{-r}^{0} \boldsymbol{y}^{\top}(t+\tau) \Big[ rS(r) + (\tau+r)U(r) \Big] \boldsymbol{y}(t+\tau)\mathsf{d}\tau, \quad (21)$$

where $L_d(\tau) = \boldsymbol{\ell}_d(\tau) \otimes I_\nu$ and $\boldsymbol{\ell}_d(\tau)$ is given in (2). Furthermore, the matrix parameters are $P(\cdot) \in \mathbb{S}^{n+\varrho}[\mathbb{R}]_{\lambda_1}$, $S(\cdot) \in \mathbb{S}^{\nu}[\mathbb{R}]_{\lambda_2}$ and $U(\cdot) \in \mathbb{S}^{\nu}[\mathbb{R}]_{\lambda_3}$ with the degree indexes $\lambda_1; \lambda_2; \lambda_3 \in \mathbb{N}_0$. It is easy to see that $\mathbb{S}^{\nu}[\mathbb{R}]_0 = \mathbb{S}^{\nu}$. Finally, it is obvious that for all $r \in [r_1, r_2]$, $v(r, \boldsymbol{0}, \boldsymbol{0}(\cdot)) = 0$ with given $r_2 > r_1 > 0$.

**Remark 5.** Note that the structure of (21) is inspired by the complete Krasovskii functional proposed in Gu & Liu (2009). Because all the matrix parameters in (21) are related to $r$ polynomially, thus it might be anticipated that less conservative results, in terms of range delay stability analysis, can be produced by (21) in comparison to a Krasovskii functional with only constant matrix parameters.

### 4.2. Optimization conditions of range dissipative stability analysis

In the following theorem, we propose optimization conditions which infer the stability and dissipativity of (1). The constraints are solved via the method of sum of squares programming.

**Theorem 1.** *Given $J_1 \prec 0$ in (13), the linear CDDS (1) is globally uniformly asymptotically stable at its origin and dissipative with respect to (13) for all $r \in [r_1, r_2]$, if there exist $P(\cdot); \widehat{P}(\cdot) \in \mathbb{S}^{n+\varrho}[\mathbb{R}]$ and $S(\cdot); \widehat{S}(\cdot); U(\cdot); \widehat{U}(\cdot) \in \mathbb{S}^{\nu}[\mathbb{R}]$ such that the following conditions hold,*

$$P(\cdot) + \left[ \mathsf{O}_n \oplus (\mathsf{D}_d \otimes S(\cdot)) \right] + g(\cdot)\widehat{P}(\cdot) \in \boldsymbol{\Sigma}_{\delta_1}\left(\mathbb{R}\,\mathring{,}\,\mathbb{S}_{\succ 0}^{n+\varrho}\right) \ \widehat{P}(\cdot) \in \boldsymbol{\Sigma}_{\delta_2}\left(\mathbb{R}\,\mathring{,}\,\mathbb{S}_{\succeq 0}^{n+\varrho}\right) \quad (22)$$

$$S(\cdot) + g(\cdot)\widehat{S}(\cdot) \in \boldsymbol{\Sigma}_{\delta_3}\left(\mathbb{R}\,\mathring{,}\,\mathbb{S}_{\succeq 0}^{n}\right), \ \widehat{S}(\cdot) \in \boldsymbol{\Sigma}_{\delta_4}\left(\mathbb{R}\,\mathring{,}\,\mathbb{S}_{\succeq 0}^{n}\right) \quad (23)$$

$$U(\cdot) + g(\cdot)\widehat{U}(\cdot) \in \boldsymbol{\Sigma}_{\delta_5}\left(\mathbb{R}\,\mathring{,}\,\mathbb{S}_{\succeq 0}^{n}\right), \ \widehat{U}(\cdot) \in \boldsymbol{\Sigma}_{\delta_6}\left(\mathbb{R}\,\mathring{,}\,\mathbb{S}_{\succeq 0}^{n}\right) \quad (24)$$

$$-\begin{bmatrix} J_1^{-1} & \Sigma(\cdot) \\ * & \boldsymbol{\Phi}_d(\cdot) \end{bmatrix} + g(\cdot)\boldsymbol{\Psi}(\cdot) \in \boldsymbol{\Sigma}_{\delta_7}\left(\mathbb{R}\,\mathring{,}\,\mathbb{S}_{\succ 0}^{m+q+2n+\varrho}\right), \ \boldsymbol{\Psi}(\cdot) \in \boldsymbol{\Sigma}_{\delta_8}\left(\mathbb{R}\,\mathring{,}\,\mathbb{S}_{\succeq 0}^{m+q+2n+\varrho}\right) \quad (25)$$

*where $g(r) = (r-r_1)(r-r_2)$ and $\delta_7 \in \mathbb{N}$ with $\delta_i; \delta_8 \in \mathbb{N}_0$, $i = 1 \cdots 6$, and*

$$\boldsymbol{\Phi}_d(r) := \textsf{Sy}\left( \begin{bmatrix} \mathsf{O}_{q \times n} & \mathsf{O}_{q \times \varrho} \\ I_n & \mathsf{O}_{n \times \varrho} \\ \mathsf{O}_n & \mathsf{O}_{n \times \varrho} \\ \mathsf{O}_{\varrho \times n} & rI_\varrho \end{bmatrix} P(r) \begin{bmatrix} D_1 & A_1 & A_2 & rA_3(r) \\ \mathsf{O}_{\varrho \times q} & L_d(0)A_4 & L_d(0)A_5 - L_d(-r) & -\widehat{\mathsf{L}}_d \end{bmatrix} \right)$$
$$+ \Gamma^{\top}\left( rS(r) + rU(r) \right)\Gamma - \left[ J_3 \oplus \mathsf{O}_n \oplus rS(r) \oplus (r\mathsf{D}_d \otimes U(r)) \right] - \textsf{Sy}\left( \begin{bmatrix} \Sigma^{\top} J_2 & \mathsf{O}_{(n+\nu+\varrho+q) \times (n+\nu+\varrho)} \end{bmatrix} \right), \quad (26)$$

$$\Gamma := \begin{bmatrix} \mathsf{O}_{\nu \times q} & A_4 & A_5 & \mathsf{O}_{\nu \times \varrho} \end{bmatrix}, \quad \Sigma(r) := \begin{bmatrix} D_2 & C_1 & C_2 & rC_3(r) \end{bmatrix}. \quad (27)$$

*Proof.* First of all, we will demonstrate that the feasible solutions of (24)–(25) *infer the existence of* (21) *satisfying* (12) *and* (10) *with* (11). Differentiating $v(r, \boldsymbol{x}(t), \boldsymbol{y}(t + \cdot))$ alongside the trajectory of (1) and considering the relation

$$\int_{-r}^{0} L_d(\tau)\dot{\boldsymbol{y}}(t+\tau)\mathsf{d}\tau = L_d(0)\boldsymbol{y}(t) - L_d(-r)\boldsymbol{y}(t-r) - \widehat{\mathsf{L}}_d \frac{1}{r}\int_{-r}^{0} L_d(\tau)\boldsymbol{y}(t+\tau)\mathsf{d}\tau$$

$$= L_d(0)A_4\boldsymbol{x}(t) + (L_d(0)A_5 - L_d(-r))\boldsymbol{y}(t-r) - \widehat{\mathsf{L}}_d \frac{1}{r}\int_{-r}^{0} L_d(\tau)\boldsymbol{y}(t+\tau)\mathsf{d}\tau \quad (28)$$

produces

$$\begin{aligned}
&\dot{v}(r, \boldsymbol{x}(t), \boldsymbol{y}(t+\cdot)) - s(\boldsymbol{z}(t), \boldsymbol{w}(t)) \\
&= \boldsymbol{\chi}_d^\top(t)\, \mathsf{Sy}\left(\begin{bmatrix} \mathsf{O}_{q\times n} & \mathsf{O}_{q\times \varrho} \\ I_n & \mathsf{O}_{n\times \varrho} \\ \mathsf{O}_{\nu\times n} & \mathsf{O}_{\nu\times \varrho} \\ \mathsf{O}_{\varrho\times n} & rI_\varrho \end{bmatrix} P(r) \begin{bmatrix} D_1 & A_1 & A_2 & rA_3(r) \\ \mathsf{O}_{\varrho\times q} & L_d(0)A_4 & L_d(0)A_5 - L_d(-r) & -\widehat{\mathsf{L}}_d \end{bmatrix}\right)\boldsymbol{\chi}_d(t) \\
&\quad + \boldsymbol{\chi}_d^\top(t)\left[\Gamma^\top\left(rS(r) + rU(r)\right)\Gamma - \left(J_3 \oplus \mathsf{O}_n \oplus rS(r) \oplus \mathsf{O}_\varrho\right)\right]\boldsymbol{\chi}_d(t) \\
&\quad - \boldsymbol{\chi}_d^\top(t)\left(\Sigma^\top(r)J_1\Sigma(r) + \mathsf{Sy}\left(\begin{bmatrix} \Sigma^\top(r)J_2 & \mathsf{O}_{(n+\nu+\varrho+q)\times(n+\nu+\varrho)} \end{bmatrix}\right)\right)\boldsymbol{\chi}_d(t), \\
&\quad - \int_{-r}^{0} \boldsymbol{y}^\top(t+\tau)U(r)\boldsymbol{y}(t+\tau)\mathsf{d}\tau,
\end{aligned} \quad (29)$$

where $\Gamma$ and $\Sigma(r)$ have been defined in (27) and $\widehat{\mathsf{L}}_d := \mathsf{L}_d' \otimes I_\nu$ in (28) can be obtained by (5) with (14). Finally,

$$\boldsymbol{\chi}_d(t) := \mathsf{Col}\left(\boldsymbol{w}(t),\ \boldsymbol{x}(t),\ \boldsymbol{y}(t-r),\ \frac{1}{r}\int_{-r}^{0} L_d(\tau)\boldsymbol{y}(t+\tau)\mathsf{d}\tau\right). \quad (30)$$

Assume $U(r) \succeq 0$, $\forall r \in [r_1, r_2]$. Considering the fact that $\boldsymbol{y}(t + \cdot) \in \widehat{\mathbb{C}}([-r, 0]\,\mathring{,}\, \mathbb{R}^\nu) \subset \mathbb{L}^2([-r, 0]\,\mathring{,}\, \mathbb{R}^\nu)$, now apply (15) to the integral $\int_{-r}^{0} \boldsymbol{y}^\top(t+\tau)U(r)\boldsymbol{y}(t+\tau)\mathsf{d}\tau$ in (29). It produces

$$\forall r \in [r_1, r_2],\quad \int_{-r}^{0} \boldsymbol{y}^\top(t+\tau)U(r)\boldsymbol{y}(t+\tau)\mathsf{d}\tau \geq [*]\left(r\mathsf{D}_d \otimes U(r)\right)\left[\int_{-r}^{0} r^{-1}L_d(\tau)\boldsymbol{y}(t+\tau)\mathsf{d}\tau\right] \quad (31)$$

where $\mathsf{D}_d = \bigoplus_{i=0}^{d} 2i+1$. Moreover, applying (31) to (29) yields

$$\forall r \in [r_1, r_2],\ \forall t \in \mathbb{T},\ \dot{v}(r, \boldsymbol{x}(t), \boldsymbol{y}(t+\cdot)) - s(\boldsymbol{z}(t), \boldsymbol{w}(t)) \leq \boldsymbol{\chi}_d^\top(t)\left(\boldsymbol{\Phi}_d(r) - \Sigma^\top(r)J_1\Sigma(r)\right)\boldsymbol{\chi}_d(t), \quad (32)$$

where $\boldsymbol{\Phi}_d(r)$ and $\boldsymbol{\chi}_d(t)$ have been defined in (26) and (30), respectively. Based on the structure of (32), it is easy to see that if

$$\forall r \in [r_1, r_2]:\quad \boldsymbol{\Phi}_d(r) - \Sigma^\top(r)J_1\Sigma(r) \prec 0,\ \ U(r) \succeq 0 \quad (33)$$

is satisfied then the dissipative inequality in (12) : $\dot{v}(r, \boldsymbol{x}(t), \boldsymbol{y}(t+\cdot)) - s(\boldsymbol{z}(t), \boldsymbol{w}(t)) \leq 0$ holds $\forall r \in [r_1, r_2]$ and $\forall t \in \mathbb{T}$.

Furthermore, by considering the fact that $J_1 \prec 0$ and the structure of $\boldsymbol{\Phi}_d(r) - \Sigma^\top(r)J_1\Sigma(r) \prec 0$, $\forall r \in [r_1, r_2]$ with the properties of negative definite matrices, it is obvious that given (33) holds then there exists (21) and $\epsilon_3 > 0$ satisfying $\forall r \in [r_1, r_2]$, $\forall t \in \mathbb{T}$, $\dot{v}(r, \boldsymbol{x}(t), \boldsymbol{y}(t+\cdot)) \leq -\epsilon_3\|\boldsymbol{x}(t)\|_2^2$. Considering (11), it shows that the feasible solutions of (33) infer the existence of $\epsilon_3 > 0$ and (21) satisfying (10). On the other hand, given $J_1 \prec 0$, applying Schur complement to (33) enables one to conclude that (33) holds if and only if

$$\forall r \in \mathcal{G}:\quad \boldsymbol{\Theta}_d(r) = \begin{bmatrix} J_1^{-1} & \Sigma(r) \\ * & \boldsymbol{\Phi}_d(r) \end{bmatrix} \prec 0,\ \ U(r) \succeq 0 \quad (34)$$

where $\mathcal{G} := \{\rho \in \mathbb{R} : g(\rho) := (\rho - r_1)(\rho - r_2) \leq 0\} = [r_1, r_2]$. Now apply the matrix sum of square relaxation technique proposed in Scherer & Hol (2006) to (34), given the fact that $g(\cdot)$ naturally satisfies the qualification constraint in the Theorem 1 of Scherer & Hol (2006). Then we can conclude that (34) holds if and only if[2] (24) and (25) hold for some $\delta_i$, $i = 5 \cdots 8$. This shows that the feasible solutions of (24)–(25) infer the existence of (21) satisfying (12) and (10) with (11).

*Now we will start to prove that the feasible solutions of* (22)–(24) *infer the existence of* (21) *satisfying* (9). Given the structure of (21), it follows that

$$\exists \lambda > 0 : \forall r \in [r_1, r_2], \forall t \in \mathbb{T}, v(r, \boldsymbol{x}(t), \boldsymbol{y}(t + \cdot)) \leq [*]\lambda \begin{bmatrix} \boldsymbol{x}(t) \\ \int_{-r}^{0} L_d(\tau)\boldsymbol{y}(t+\tau)\mathrm{d}\tau \end{bmatrix} + \int_{-r}^{0} [*]\lambda \boldsymbol{y}(t+\tau)\mathrm{d}\tau$$

$$\leq \lambda \|\boldsymbol{x}(t)\|_2^2 + [*]\lambda \int_{-r}^{0} L_d(\tau)\boldsymbol{y}(t+\tau)\mathrm{d}\tau + \lambda r\|\boldsymbol{y}(t+\cdot)\|_\infty^2 \leq \lambda \|\boldsymbol{x}(t)\|_2^2 + \lambda r\|\boldsymbol{y}(t+\cdot)\|_\infty^2$$

$$+ [*]\left(\lambda \mathsf{D}_d \otimes I_n\right) \int_{-r}^{0} L_d(\tau)\boldsymbol{y}(t+\tau)\mathrm{d}\tau \leq \lambda\|\boldsymbol{x}(t)\|_2^2 + \lambda r\|\boldsymbol{y}(t+\cdot)\|_\infty^2 + r\int_{-r}^{0} [*]\lambda\boldsymbol{y}(t+\tau)\mathrm{d}\tau$$

$$\leq \lambda\|\boldsymbol{x}(t)\|_2^2 + \left(\lambda r + \lambda r^2\right)\|\boldsymbol{y}(t+\cdot)\|_\infty^2 \leq \left(\lambda + \lambda r^2\right)\|\boldsymbol{x}(t)\|_2^2 + \left(\lambda r + \lambda r^2\right)\|\boldsymbol{y}(t+\cdot)\|_\infty^2$$

$$\leq 2\left(\lambda r + \lambda r^2\right)(\|\boldsymbol{x}(t)\|_2 \vee \|\boldsymbol{y}(t+\cdot)\|_\infty)^2 \leq 2\left(\lambda r + \lambda r_2^2\right)(\|\boldsymbol{x}(t)\|_2 \vee \|\boldsymbol{y}(t+\cdot)\|_\infty)^2. \quad (35)$$

Consequently, it shows that there exist $\epsilon_2 > 0$ and a functional with the form of (21) satisfying $\forall r \in [r_1, r_2]$, $\forall t \in \mathbb{T}$, $v(r, \boldsymbol{x}(t), \boldsymbol{y}(t+\cdot)) \leq \epsilon_2 (\boldsymbol{x}(t) \vee \|\boldsymbol{y}(t+\cdot)\|_\infty)^2$.

Now assume $S(r) \succeq 0, \forall r \in [r_1, r_2]$. Then applying (15) to the integral $\int_{-r}^{0} \boldsymbol{y}^\top(t+\tau)S(r)\boldsymbol{y}(t+\tau)\mathrm{d}\tau$ in (21) yields

$$\forall r \in [r_1, r_2], \quad r\int_{-r}^{0} \boldsymbol{y}^\top(t+\tau)S(r)\boldsymbol{y}(t+\tau)\mathrm{d}\tau \geq [*]\left(\mathsf{D}_d \otimes S(r)\right)\int_{-r}^{0} L_d(\tau)\boldsymbol{y}(t+\tau)\mathrm{d}\tau, \quad (36)$$

where $\mathsf{D}_d = \bigoplus_{i=0}^{d} 2i+1$. Applying (36) to (21) produces that $\forall r \in [r_1, r_2], \forall t \in \mathbb{T}$

$$v(r, \boldsymbol{x}(t), \boldsymbol{y}(t+\cdot)) \geq [*]\left(P(r) + \begin{bmatrix} \mathsf{O}_n \oplus (\mathsf{D}_d \otimes S(r)) \end{bmatrix}\right)\begin{bmatrix} \boldsymbol{x}(t) \\ \int_{-r}^{0} L_d(\tau)\boldsymbol{y}(t+\tau)\mathrm{d}\tau \end{bmatrix}$$

$$+ \int_{-r}^{0} (\tau + r)\boldsymbol{y}^\top(t+\tau)U(r)\boldsymbol{y}(t+\tau)\mathrm{d}\tau. \quad (37)$$

Considering the structure of (37), it is obvious to see that if

$$\forall r \in \mathcal{G} : \quad \Pi_d(r) := P(r) + \begin{bmatrix} \mathsf{O}_n \oplus (\mathsf{D}_d \otimes S(r)) \end{bmatrix} \succ 0, \quad S(r) \succeq 0, \quad U(r) \succeq 0 \quad (38)$$

is satisfied, then $\exists \epsilon_1 > 0: \forall r \in [r_1, r_2], \forall t \in \mathbb{T}, v(r, \boldsymbol{x}(t), \boldsymbol{y}(t+\cdot)) \geq \epsilon_1 \|\boldsymbol{x}(t)\|_2$, where $\mathcal{G} := \{\rho \in \mathbb{R} : g(\rho) := (\rho - r_1)(\rho - r_2) \leq 0\} = [r_1, r_2]$. Now apply the matrix relaxation technique in Scherer & Hol (2006) to the conditions in (38). Then one can conclude that (38) holds if and only if (22)–(24) hold for some $\delta_i$, $i = 1 \cdots 6$. Considering the upper bound result which has been derived in (35), one can see the feasible solutions of (22)–(24) infer the existence of (21) and $\epsilon_1; \epsilon_2 > 0$ satisfying $\forall r \in [r_1, r_2], \forall t \in \mathbb{T}$, $\epsilon_1 \|\boldsymbol{x}(t)\|_2 \leq v(r, \boldsymbol{x}(t), \boldsymbol{y}(t+\cdot)) \leq \epsilon_2 (\boldsymbol{x}(t) \vee \|\boldsymbol{y}(t+\cdot)\|_\infty)^2$. Finally, by considering (11), we have proved that the feasible solutions of (22)–(24) infer the existence of (21) and $\epsilon_1; \epsilon_2 > 0$ satisfying (9).

In conclusion, we have demonstrated that the feasible solutions of (22)–(25) infer the existence of $\epsilon_1; \epsilon_2; \epsilon_3 > 0$ and (21) satisfying (9), (10) considering (11), and (12). This finishes the proof. $\square$

---

[2] See the results related to the equations (1) and (6) in Scherer & Hol (2006)

**Remark 6.** All SoS constraints in Theorem 1 can be solved numerically via the relation in (20). The dimension of the corresponding certificate variable $Q$ in (20) is determined by the values of $\delta_i$, $i = 1 \cdots 8$ with $\lambda_1$, $\lambda_2$ and $\lambda_3$ in (21). Meanwhile, if $J_1 \preceq 0$ and $J_1 \not\prec 0$, the corresponding Theorem 1 can be easily derived based on the factorization illustrated in Scherer *et al.* (1997). Furthermore, Schur complement does not need to be applied to $\Sigma^\top J_1 \Sigma$ if $J_1 = \mathsf{O}_m$, which corresponds to the modeling of strict passivity constraint.

**Remark 7.** One can use different forms of $g(r)$ to characterize the set $\mathcal{G} = [r_1, r_2] = \{r \in \mathbb{R} : g(r) \leq 0\}$. Note that the form $g(r) = (r - r_1)(r - r_2)$ in (34) is one of many valid examples which can equivalently transfer the conditions in (34) and (38) into SoS constraints, as long as $g(r) \leq 0$ can equivalently characterize the interval $[r_1, r_2]$ and satisfy the qualification constraints in Scherer & Hol (2006). This also infers that valid $g(r)$ with different forms do not bring changes to the feasibility of the corresponding SoS constraints, since they ultimately are equivalent to (34) and (38). Nevertheless, the form $g(r) = (r - r_1)(r - r_2)$ might be the best option to solve (34) and (38) considering its low degree, which alleviates the computational burden to solve (22)–(25).

**Remark 8.** Point-wise delay stability analysis at $r = r_0 > 0$ can be achieved by solving

$$S \succeq 0,\ U \succeq 0,\ \Pi \succ 0,\ \mathbf{\Theta}_d(r_0) \prec 0, \tag{39}$$

in which the value of $r_0$ is given and $\lambda_1 = \lambda_2 = \lambda_3 = 0$ in (21). Since $r_0$ here is of fixed value, a non-constant polynomials matrix variables in terms of $r$ in (21) will be meaningless. We emphasize that every time when (39) is cited in this paper, it is in line with the assumption $\lambda_1 = \lambda_2 = \lambda_3 = 0$.

*4.3. Reducing the computational burden of Theorem 1 for certain cases*

The SoS constraints in (22)–(25) can be applied to any form of (1) with given values of $\lambda_1$, $\lambda_2$, $\lambda_3$ in (21), supported by proper choice of $\delta_i$, $i = 1 \cdots 8$. However, if any inequality in (34) or (38) is affine with respect to $r$, then it can be solved equivalently via the property of convex hull to save unnecessary variables instead of solving the equivalent SoS constraints in (22)–(25). Nevertheless, this can only happen to very special cases as what are summarized as follows.

**Case 1.** If $\lambda_1 = \lambda_2 = \lambda_3 = 0$ and $A_3(r) = 0$, then (22)–(24) become standard LMIs by letting $\delta_i = 0$, $i = 1 \cdots 6$ and $\widehat{P}(r) = \mathsf{O}_{\nu+\varrho}$, $\widehat{S}(r) = \widehat{U}(r) = \mathsf{O}_\nu$, which in this case directly corresponds to the LMIs in (38). On the other hand, the matrix inequality $\mathbf{\Theta}_d(r) \prec 0$ in (34) is affine with respect to $r$. Thus $\forall r \in [r_1, r_2]$, $\mathbf{\Theta}_d(r) \prec 0$ can be solved by the property of convex hull instead of using (25). Meanwhile, $\forall r \in [r_1, r_2]$, $\mathbf{\Theta}_d(r) \prec 0$ here can still be solved via the SoS condition (25), with the degrees $\delta_7 = 1$ and $\delta_8 = 0$ for example. However, since using SoS does not add any extra feasibility, it is preferable in this case to use the property of convex hull to reduce the number of decision variables.

**Case 2.** If any inequality in (38) is affine (convex)[3] with respect to $r$, then it can be solved directly via the property of convex hull. Meanwhile, if unstructured matrix variables are considered in (21) without predefined sparsities, then the only possibility for $\mathbf{\Theta}_d(r) \prec 0$ in (34) to be an affine (convex) matrix inequality in $r$ is when $\lambda_1 = \lambda_2 = \lambda_3 = 0$ and $A_3(r) = 0$ as what has been stated in Case 1. (see the Remark 5 in Gyurkovics & Takács (2016)) Therefore, the property of convex hull cannot be applied to solve the parameter dependent inequality $\forall r \in [r_1, r_2]$, $\mathbf{\Theta}_d(r) \prec 0$ if (1) contains a distributed delay term.

We have demonstrated that for certain situations, one can solve (38) and (34) via the property of convex hull with less number of decision variables instead of solving (22)–(25). However, based on the elaborations we have made in Case 1 and Case 2, the SoS constraint in (25) is necessary to be applied if (1) has non-zero distributed delay terms, which is still true even if (21) has only constant matrix parameters ($\lambda_1 = \lambda_2 = \lambda_3 = 0$).

---

[3]This may include the situation such as $S(r) = S_1 + r^4 S_2$. However, the handling of $\forall r \in [r_1, r_2]$, $S(r) \succ 0$ is identical to an affine example. In addition, the variable structure such as $S(r) = r^3 S_1$ will not be considered in this paper, since always it can be equivalently transferred into a constant parameter.

*4.4. Estimating delay margins subject to prescribed performance objectives*

Given an initial $r_0$ together with a prescribed performance objective (no decision variables in (13)) which renders (39) to be feasible, we are interested in the following problem.

**Problem 1.** Finding the minimum $\grave{r}$ or maximum $\acute{r}$ which render (1) to be stable and dissipative over $[\grave{r}, r_0]$ or $[r_0, \acute{r}]$, where the dissipative constraint in (13) is given and it is satisfied at $r_0$.

The control interpretation of this problem is straightforward: Given a specific performance objective, we want to obtain the largest stable delay interval of a delay system over which the system can always satisfy the given performance objective.

Problem (1) can be solved by the following optimization programs

$$\min \rho \quad \text{subject to} \quad (22)-(25) \quad \text{with} \quad g(r) = (r-\rho)(r-r_0) \tag{40}$$

or

$$\max \rho \quad \text{subject to} \quad (22)-(25) \quad \text{with} \quad g(r) = (r-r_0)(r-\rho), \tag{41}$$

with given $\lambda_1, \lambda_2, \lambda_3$ and $\delta_i$, $i = 1 \cdots 8$ values. Specifically, (40) and (41) can be easily handled via an iterative one dimensional search scheme together with the idea of bisections Boyd & Vandenberghe (2004). Since (40) and (41) are all of range stability conditions, thus the usage of bisections will not produce false feasible solutions even (40) and (41) are not necessarily quasi-convex. Furthermore, as what have been elaborated in subsection 4.3, if any inequality in (38) and (34) is affine (convex) with respect to $r$, then it can be solved directly for (40) and (41) via the property of convex hull instead of solving the corresponding SoS conditions in (22)-(25). Finally, It is very important to emphasize here that the dissipative results over $[\grave{r}, r_0]$ or $[r_0, \acute{r}]$, produced individually by (40) and (41), cannot be automatically merged together due to the nature of dissipative analysis.

*4.5. Hierarchical stability result*

Here we show that the range stability conditions in Theorem 1 exhibit a hierarchical enhancement with respect to $d$.

**Theorem 2.** *Given the same prerequisites in Theorem 1, we have*

$$\forall d \in \mathbb{N}_0, \quad \mathcal{F}_d \subseteq \mathcal{F}_{d+1} \tag{42}$$

*where*

$$\begin{aligned}
\mathcal{F}_d &:= \Big\{ (r_1, r_2) \,\Big|\, r_2 > r_1 > 0 \quad \& \quad (38) \text{ and } (34) \text{ hold} \Big\} \\
&= \Big\{ (r_1, r_2) \,\Big|\, r_2 > r_1 > 0 \quad \& \quad (22)\text{--}(25) \text{ hold} \quad \& \quad \delta_7 \in \mathbb{N} \quad \& \quad \delta_i; \delta_8 \in \mathbb{N}_0, i = 1 \cdots 6 \Big\}.
\end{aligned} \tag{43}$$

*Proof.* Refers to Appendix A. □

## 5. Numerical Examples

All numerical examples in this section are calculated in Matlab environment using Yalmip Löfberg (2004) as the optimization interface. In addition, Mosek (2016) is applied as the SDP numerical solvers. Moreover, all SoS constraints are implemented via the function `coefficient` in Yalmip. Finally, we stress that no prescribed positive eigenvalue margins are imposed for PSD (positive semi-definite) variables in our programs. Instead, a valid feasible result (optimization programs follow the same principle) must be the case that all eigenvalues of the resulting PSD variables[4] are strictly positive, regardless of what other parameters might be reported by the numerical solvers.

---

[4] This includes all the SoS certificate variables.

*5.1. Delay range stability analysis*

In this subsection, we consider analyzing the range stability of

$$\dot{\boldsymbol{x}}(t) = A_1\boldsymbol{x}(t) + A_2\boldsymbol{y}(t-r) + \int_{-r}^{0} A_3(\tau)\boldsymbol{y}(t+\tau)\mathsf{d}\tau \qquad (44)$$
$$\boldsymbol{y}(t) = A_4\boldsymbol{x}(t) + A_5\boldsymbol{y}(t-r)$$

with different state space parameters presented in Table 1, in which the delay margins $r_{\min}$ and $r_{\max}$ are calculated via the software package Breda *et al.* (2014) in Matlab with reference to the spectral method in Breda *et al.* (2005).

| Parameters | $A_1$ | $A_2$ | $A_3(r)L_d(\tau)$ | $A_4$ | $A_5$ | $r_{\min}$ | $r_{\max}$ |
|---|---|---|---|---|---|---|---|
| Example 1 | $\begin{bmatrix} 0 & 1 \\ -2 & 0.1 \end{bmatrix}$ | $\begin{bmatrix} 0 \\ 1 \end{bmatrix}$ | $\begin{bmatrix} 0 \\ 0 \end{bmatrix}$ | $\begin{bmatrix} 0 & 1 \end{bmatrix}$ | $0$ | 0.10016827 | 1.71785 |
| Examples 2 | $\begin{bmatrix} 0.2 & 0.01 \\ 0 & -2 \end{bmatrix}$ | $\mathsf{O}_2$ | $\begin{bmatrix} -1+0.3\tau & 0.1 \\ 0 & -0.1 \end{bmatrix}$ | $\begin{bmatrix} 1 & 0 \\ 0 & 1 \end{bmatrix}$ | $\mathsf{O}_2$ | 0.1944 | 1.7145 |

Table 1: Numerical Examples of (44)

The examples in Table 1 are taken from Ariba *et al.* (2017) and Gouaisbaut & Ariba (2011), respectively, which are denoted via the equivalent CDDS representations. To the best of our knowledge, Gouaisbaut & Ariba (2011) is the only published paper to be able to handle range stability analysis for systems having a distributed delay with polynomials kernels. Meanwhile, Example 2 cannot be analyzed by the range stability results in Gyurkovics & Takács (2016); Seuret & Gouaisbaut (2015); Gouaisbaut *et al.* (2013).

Note that the matrix $A_3(r)$ in Table 1 corresponding to Example 1 and Example 2 are $A_3(r) = \mathsf{O}_{2\times(d+1)}$ and

$$A_3(r) = \begin{bmatrix} -1 & 0.1 & 0.3 & 0 & \mathsf{O}_{2\times(2d-2)} \\ 0 & -0.1 & 0 & 0 & \end{bmatrix} (\mathsf{L}_d^{-1}(r) \otimes I_2) = \begin{bmatrix} -1-0.15r & 0.1 & 0.15r & 0 & \mathsf{O}_{2\times(2d-2)} \\ 0 & -0.1 & 0 & 0 & \end{bmatrix}, \qquad (45)$$

respectively. In the following Table 2–3, the results of maximum detectable stable delay interval calculated by our method are presented compared to the results in Ariba *et al.* (2017) and Gouaisbaut & Ariba (2011), respectively. Note that the values of $\delta_7$ and $\delta_8$ therein are the degrees of the SoS constraint (25). In addition, as what we have stated in subsection 4.4 concerning the reduction of the numerical complexity of Theorem 1, if any inequality in (38) is affine, then it is solved via the property of convex hull to reduce computational burdens.

| Solutions for Delay Range Stability | $[r_1, r_2]$ | NoV |
|---|---|---|
| Ariba *et al.* (2017) ($N=5$) | $[0.10016829, 1.7178]$ | 294 |
| Theorem 1 ($\lambda_1=1, \lambda_2=\lambda_3=0, d=4, \delta_7=1, \delta_8=0$) | $[0.10016828, 1.71785]$ | 231 |
| Theorem 1 ($\lambda_1=1, \lambda_2=\lambda_3=0, d=5, \delta_7=1, \delta_8=0$) | $[0.10016827, 1.71785]$ | 291 |

Table 2: Largest detectable stable interval of Example 1. The analytic stable interval is $[0.10016827, 1.71785]$

| Solutions for Delay Range Stability | $[r_1, r_2]$ | NoV |
|---|---|---|
| Gouaisbaut & Ariba (2011) ($l=1, r=3$) | $[0.2, 1.29]$ | 5973 |
| Gouaisbaut & Ariba (2011) ($l=2, r=3$) | $[0.2, 1.3]$ | 14034 |
| Theorem 1 ($\lambda_1=\lambda_2=\lambda_3=0, d=4, \delta_7=2, \delta_8=1$) | $[0.27, 1.629]$ | 1394 |
| Theorem 1 ($\lambda_1=1, \lambda_2=\lambda_3=0, d=4, \delta_7=2, \delta_8=1$) | $[0.1944, 1.7145]$ | 1472 |

Table 3: Largest detectable stable interval of Example 2. The analytic stable interval is $[0.1944, 1.7145]$.

From the results summarized in Table 2-3, one can clearly observe the advantage of our proposed methods as the stable intervals of Example 1 and 2, in line with the delay margins calculated by the method in Breda *et al.* (2014), can be detected with less variables compared to the existing results in Gouaisbaut & Ariba (2011) and Ariba *et al.* (2017). In addition, one can clearly see from table 3 the benefits in applying (21) with delay-dependent matrix parameters. In addition, note that our method does not require the constraint $A + A_d$ being nonsingular as the Theorem 4 of Ariba *et al.* (2017) does.

**Remark 9.** Note that the number of decision variables of Theorem 1 in Table 2–3 might be further reduced by simplifying the SoS certificate variable in (20) for each case when a SoS condition needs to be solved.

*5.2. Dissipative Range Stability Analysis*

Consider the following neutral delay system

$$\frac{\mathrm{d}}{\mathrm{d}t}\left(\boldsymbol{y}(t) - A_4\boldsymbol{y}(t-r)\right) = A_1\boldsymbol{y}(t) + A_2\boldsymbol{y}(t-r) + \int_{-r}^{0} A_3(r)L_d(\tau)\boldsymbol{y}(t+\tau)\mathrm{d}\tau + D_1\boldsymbol{w}(t)$$
$$\boldsymbol{z}(t) = C_1\boldsymbol{y}(t) + C_2\boldsymbol{y}(t-r) + \int_{-r}^{0} C_3(r)L_d(\tau)\boldsymbol{y}(t+\tau) + D_2\boldsymbol{w}(t)$$
(46)

with distributed delay terms at the state and output. Let $\boldsymbol{x}(t) = \boldsymbol{y}(t) - A_4\boldsymbol{y}(t-r)$, then (46) can be reformulated into

$$\dot{\boldsymbol{x}}(t) = A_1\boldsymbol{x}(t) + (A_1A_4 + A_2)\boldsymbol{y}(t-r) + \int_{-r}^{0} A_3(r)L_d(\tau)\boldsymbol{y}(t+\tau) + D_1\boldsymbol{w}(t),$$
$$\boldsymbol{y}(t) = \boldsymbol{x}(t) + A_4\boldsymbol{y}(t-r),$$
$$\boldsymbol{z}(t) = C_1\boldsymbol{x}(t) + (C_1A_3 + C_2)\boldsymbol{y}(t-r) + \int_{-r}^{0} C_3(r)L_d(\tau)\boldsymbol{y}(t+\tau) + D_2\boldsymbol{w}(t).$$
(47)

which is now in line with the CDDS form in (1). Now consider a linear neutral delay system with the parameters $A_3L_d(\tau) = C_3L_d(\tau) = \mathsf{O}_3$ and

$$A_1 = 100\begin{bmatrix} -2.103 & 1 & 2 \\ 3 & -9 & 0 \\ 1 & 2 & -6 \end{bmatrix}, \ A_2 = 100\begin{bmatrix} 1 & 0 & -3 \\ -0.5 & -0.5 & -1 \\ -0.5 & -1.5 & 0 \end{bmatrix}, \ A_4 = \frac{1}{72}\begin{bmatrix} -1 & 5 & 2 \\ 4 & 0 & 3 \\ -2 & 4 & 1 \end{bmatrix}$$
$$D_1 = \begin{bmatrix} 0 \\ 0 \\ 0.1 \end{bmatrix}, \ C_1 = \begin{bmatrix} -0.1 & 0.1 & 0.2 \\ 0.4 & 0.01 & 0 \\ 0.1 & 0.21 & 0.1 \end{bmatrix}, \ C_2 = \begin{bmatrix} 0.1 & 0 & 0.2 \\ 0.4 & 0 & -0.1 \\ 0 & -0.5 & 0.3 \end{bmatrix}, \ D_2 = \begin{bmatrix} 0 \\ 0.1 \\ 0 \end{bmatrix}$$
(48)

which is modified based on the circuit model in Zhang (2009). We consider $\mathbb{L}^2$ gain $\gamma$ as the performance objective for (48).

Apply Theorem 1 to (47) with the parameters in (48) and $A_3L_d(\tau) = C_3L_d(\tau) = \mathsf{O}_3$. In addition, we assume $\lambda_1 = 1$ and $\lambda_2 = \lambda_3 = 0$ in (21) with a delay range $[r_1, r_2] = [0.1, 0.5]$. Since all inequalities in (38) in this case are either affine with respect to $r$ or simple LMIs, then they will be solved directly via the property of convex hull as we have discussed in subsection 4.3. The results of $\min \gamma$ over $[0.1, 0.5]$ are summarized in Table 4. Note that $\delta_7$ and $\delta_8$ are the degrees of the SoS constraints in (25).

| Theorem 1 | $\delta_7 = 1, \delta_8 = 0$ | $\delta_7 = 2, \delta_8 = 1$ | $\delta_7 = 3, \delta_8 = 2$ |
|:---:|:---:|:---:|:---:|
| $d = 1$ | 0.441 | 0.441 | 0.441 |
| $d = 2$ | 0.364 | 0.364 | 0.364 |
| $d = 3$ | 0.361 | 0.361 | 0.361 |

Table 4: values of $\min \gamma$ valid over $[0.1, 0.5]$

Now apply Theorem 1 with constant matrix parameters $\lambda_1 = \lambda_2 = \lambda_3 = 0$ in (21) to the same aforementioned model with $[r_1, r_2] = [0.1, 0.5]$. Again the conditions in (38) are simple LMIs and $\mathbf{\Theta}_d(r) \prec 0$ in (34) can be solved by the property of convex hull. However, even with $d = 10$, the range dissipative conditions here with $\lambda_1 = \lambda_2 = \lambda_3 = 0$ still cannot yield feasible solutions. This demonstrates the advantage and necessity to apply a Krasovskii functional with delay dependent parameters when a functional with constant parameters is simply not strong enough to handle a stable delay interval.

To partially verify the results in Table 4, we apply the `sigma` function[5] in Matlab, which can calculate the singular values ($\min \gamma$) of a dynamical system over a fixed frequency range. By extracting the peak value produced by `sigma`, it yields that the system (47) with (48) and $A_3 L_d(\tau) = C_3 L_d(\tau) = \mathbf{O}_3$ produces $\min \gamma = 0.101074$ and $\min \gamma = 0.36064$ at $r = 0.1$ and $r = 0.5$, respectively. This shows that the values of $\min \gamma$ in Table 4, which are valid over $[0.1, 0.5]$, are consistent with the point-wise $\min \gamma$ values obtained via `sigma` function. In addition, the best value $\min \gamma = 0.361$ in Table 4 is quite closed to the point-wise result $\min \gamma = 0.36064$ at $r = 0.5$.

Now consider new $A_3 L_d(\tau)$ and $C_3 L_d(\tau)$ with the structures

$$A_3(r)L_d(\tau) = \begin{bmatrix} 0.1\tau & 0.1 & 0.3 \\ 0.2 & 0.1 & 0.3 - 0.1\tau \\ -0.1 & -0.2 + 0.1\tau & 0.2 \end{bmatrix}, \quad C_3(r)L_d(\tau) = \begin{bmatrix} 0.1 & 0 & 0.2 \\ 0.4 & 0 & -1 \\ 0 & -0.5 & 0.3 \end{bmatrix}, \quad (49)$$

which together with (48) constitute a linear neutral system with distributed delays. Note that we can easily obtain the corresponding matrix coefficients as

$$A_3(r) = \begin{bmatrix} -0.05r & 0.1 & 0.3 & 0.05r & 0 & 0 \\ 0.2 & 0.1 & 0.3 + 0.05r & 0 & 0 & -0.05r & \mathbf{O}_{3\times(3d-3)} \\ -0.1 & -0.2 - 0.05r & 0.2 & 0 & 0.05r & 0 \end{bmatrix},$$

$$C_3(r) = \begin{bmatrix} 0.1 & 0 & 0.2 \\ 0.4 & 0 & -1 & \mathbf{O}_{3\times 3d} \\ 0 & -0.5 & 0.3 \end{bmatrix}. \quad (50)$$

To the best of our knowledge, there are no existing results on delay range dissipative analysis concerning linear neutral systems with non-constant distributed delay kernels. As a result, we might claim that no existing schemes can handle the problem we are dealing with here.

Now apply Theorem 1 to the system (47) with (48) and (50) and assume again $\lambda_1 = 1$ and $\lambda_2 = \lambda_3 = 0$ in (21) with the delay range $[r_1, r_2] = [0.1, 0.5]$. Once more, since all the corresponding inequalities in (38) are affine with respect to $r$, then they are directly solved via the property of convex hull instead of solving (22)–(24). The values of the resulting $\min \gamma$ over $[0.1, 0.5]$ are summarized in Table 5, where $\delta_7$ and $\delta_8$ are the degrees of the SoS constraint (25).

| Theorem 1 | $\delta_7 = 2, \delta_8 = 1$ | $\delta_7 = 3, \delta_8 = 2$ |
|---|---|---|
| $d = 1$ | 0.47 | 0.47 |
| $d = 2$ | 0.382 | 0.382 |
| $d = 3$ | 0.37822 | 0.37822 |

Table 5: values of $\min \gamma$ valid over $[0.1, 0.5]$

Unfortunately, even for the case of point-wise delays, the `sigma` function in Matlab cannot handle a distributed delay system at the current stage. Thus we suggest here to use (39) to partially verify the results in Table 5. Specifically, apply (39) with $d = 3$ to the system (47) with (48) and (50) at $r_0 = 0.1$ and $r_0 = 0.5$, respectively. It yields $\min \gamma = 0.10101$ at $r_0 = 0.1$ and $\min \gamma = 0.37822$ at $r_0 = 0.5$, respectively.

---

[5]We first use the default range of `sigma` to determine which frequency range contains the peak singular value. Based on the previous information, we assigned $\mathtt{w} = \mathtt{logspace}(-1, 2, 2000000)$ to `sigma` to ensure the accuracy of $\min \gamma$.

This verifies that the values of range $\min \gamma$ in Table 5 are consistent with the point-wise $\min \gamma$ values we presented.

Finally, the estimation problem described in subsection 4.4 can be easily applied to the system (47) with (48) and (50), given that the value of $\gamma$ in (13) is prescribed. To be specific, consider the system (47) with (48) and (50) and let $\gamma = 0.37822$ and $r_0 = 0.5$ with which feasible solutions can be produced by (39) with $d = 3$. Now, one can use (40)[6] with $\lambda_1 = 1$ and $\lambda_2 = \lambda_3 = 0$, $d = 3$ and $\delta_7 = 2$, $\delta_8 = 1$ to find out the minimum $r^*$, which renders the corresponding system to be stable and satisfying $\gamma = 0.37822$ over $[r^*, r_0]$. Given what we have presented in Table 5, it is predictable that the optimal value of $r^*$ here is $r^* = 0.1$.

## 6. Conclusion

In this paper, the solutions concerning range stability analysis for a CDDS subject to dissipative constraints have been presented. The advantage of the proposed methodologies is rooted in the application of a Krasovskii functional with delay dependent parameters, which leads to dissipative range stability conditions expressed in terms of SoS constraints. The tests of numerical examples have demonstrated that less conservative results with less computational burdens can be produced by our methods compared to existing approaches. In addition, the proposed scheme is able to handle distributed delays with polynomials kernels when dissipative range stability analysis is concerned. Meanwhile, it has been demonstrated that our approach can also handle delay margins estimation problem with prescribed performance objectives.

## 7. Acknowledgment

Many thanks to Prof. Carsten Scherer for his comments on the matrix relaxation technique in Scherer & Hol (2006). The author also thank Prof. Johan Löfberg for his help on the usage of Yalmip Löfberg (2004). Finally, the authors would like to thank the associated editor and two anonymous reviewers for their constructive comments.

## Appendix A. Proof of Theorem 2

*Proof.* Let $d \in \mathbb{N}_0$ and $(r_1, r_2) \in \mathcal{F}_d$ with $\mathcal{F}_d \neq \varnothing$. Assume that $P_{d+1}(r) := P_d(r) \oplus \mathsf{O}_n$. Given (38) with the structure of (21), we have

$$\forall r \in [r_1, r_2] : \Pi_{d+1}(r) = P_{d+1}(r) + \Big[\mathsf{O}_n \oplus (\mathsf{D}_{d+1} \otimes S(r))\Big] = P_d(r) \oplus [(2d+3) \otimes S(r)] \succ 0, \quad (\text{A.1})$$
$$\forall r \in [r_1, r_2] : \ S(r) \succeq 0, \quad U(r) \succeq 0$$

Therefore, we have shown that the existence of feasible solutions of $\forall r \in [r_1, r_2]$, $\Pi_d(r) \succ 0$ infers the existence of a feasible solution of $\forall r \in [r_1, r_2]$, $\Pi_{d+1}(r) \succ 0$, given $\forall r \in [r_1, r_2]$, $S(r) \succeq 0$, $U(r) \succeq 0$.

Given (34) with the structure of $\boldsymbol{\Theta}_d(r)$ and (30), it is obvious to see that

$$\forall r \in [r_1, r_2], \ \boldsymbol{\Theta}_{d+1}(r) = \begin{bmatrix} J_1^{-1} & \big[\Sigma(r) \ \ \mathsf{O}_{m \times n}\big] \\ * & \boldsymbol{\Phi}_{d+1}(r) \end{bmatrix} = \begin{bmatrix} J_1^{-1} & \big[\Sigma(r) \ \ \mathsf{O}_{m \times n}\big] \\ * & \boldsymbol{\Phi}_d(r) \oplus -r(2d+3)U(r) \end{bmatrix} \prec 0, \quad (\text{A.2})$$

which is derived based on the assumption $P_{d+1}(r) := P_d(r) \oplus \mathsf{O}_n$ considering the structure of $\boldsymbol{\Phi}_d(r)$ in (26). As a result, we can conclude that $\forall r \in [r_1, r_2]$, $\boldsymbol{\Theta}_d(r) \prec 0$ infers $\forall r \in [r_1, r_2]$, $\boldsymbol{\Theta}_{d+1}(r) \prec 0$ given $\forall r \in [r_1, r_2]$, $U(r) \succeq 0$. Consequently, we have shown that the existence of feasible solutions of (38) and (34) at $d$ infers the existence of the feasible solutions of (A.1) and (A.2) with $d + 1$. Finally, since (A.1) and (A.2) can be equivalently verified by the SoS conditions (22)–(25) at $d + 1$ for some $\delta_7 \in \mathbb{N}$ and $\delta_i; \delta_8 \in \mathbb{N}_0$, $i = 1 \cdots 6$, thus Theorem 2 is proved. □

---

[6] All the corresponding inequalities in (38) for (40) here are solved directly via the property of convex hull instead of solving (22)–(24).

Ariba, Yassine, Gouaisbaut, Frédéric, Seuret, Alexandre, & Peaucelle, Dimitri. 2017. Stability analysis of time-delay systems via Bessel inequality: A quadratic separation approach. *International Journal of Robust and Nonlinear Control*, n/a–n/a. rnc.3975.

Blekherman, Grigoriy, Parrilo, Pablo A, & Thomas, Rekha R. 2013. *Semidefinite optimization and convex algebraic geometry*. Vol. 13. Siam.

Boyd, Stephen, & Vandenberghe, Lieven. 2004. *Convex optimization*. Cambridge university press.

Breda, Dimitri, Maset, Stefano, & Vermiglio, Rossana. 2005. Pseudospectral differencing methods for characteristic roots of delay differential equations. *SIAM Journal on Scientific Computing*, **27**(2), 482–495.

Breda, Dimitri, Maset, Stefano, & Vermiglio, Rossana. 2014. *Stability of Linear Delay Differential Equations: A Numerical Approach with MATLAB*. Springer.

Briat, Corentin. 2014. *Linear Parameter-Varying and Time-Delay Systems*. Springer.

Chesi, G. 2010. LMI techniques for optimization over polynomials in control: A survey. *IEEE Transactions on Automatic Control*, **55**(11), 2500–2510. cited By 156.

Feng, Q., & Nguang, S. K. 2016a (Dec). Orthogonal functions based integral inequalities and their applications to time delay systems. *Pages 2314–2319 of: 2016 IEEE 55th Conference on Decision and Control (CDC)*.

Feng, Qian, & Nguang, Sing Kiong. 2016b. Stabilization of uncertain linear distributed delay systems with dissipativity constraints. *Systems & Control Letters*, **96**, 60 – 71.

Fridman, Emilia. 2014. *Introduction to Time-Delay Systems*. Springer.

Gautschi, Walter. 2004. Orthogonal polynomials: computation and approximation.

Gouaisbaut, F., & Ariba, Y. 2011. Delay range stability of a class of distributed time delay systems. *Systems & Control Letters*, **60**(3), 211 – 217.

Gouaisbaut, F., Ariba, Y., & Seuret, A. 2013 (Dec). Bessel inequality for robust stability analysis of time-delay system. *Pages 928–933 of: 52nd IEEE Conference on Decision and Control*.

Gu, K., Jin, C., Boussaada, I., & Niculescu, S. I. 2016 (Dec). Towards more general stability analysis of systems with delay-dependent coefficients. *Pages 3161–3166 of: 2016 IEEE 55th Conference on Decision and Control (CDC)*.

Gu, Keqin, & Liu, Yi. 2009. Lyapunov Krasovskii functional for uniform stability of coupled differential-functional equations. *Automatica*, **45**(3), 798 – 804.

Gu, Keqin, Han, Qing-Long, Luo, Albert CJ, & Niculescu, Silviu-Iulian. 2001. Discretized Lyapunov functional for systems with distributed delay and piecewise constant coefficients. *International Journal of Control*, **74**(7), 737–744.

Gu, Keqin, Chen, Jie, & Kharitonov, V. 2003. Stability of Time-Delay Systems.

Gumussoy, Suat, & Michiels, Wim. 2011. Fixed-Order H-Infinity Control for Interconnected Systems Using Delay Differential Algebraic Equations. *SIAM Journal on Control and Optimization*, **49**(5), 2212–2238.

Gyurkovics, É., & Takács, T. 2016. Multiple integral inequalities and stability analysis of time delay systems. *Systems & Control Letters*, **96**, 72 – 80.

Hale, Jack K, & Lunel, Sjoerd M Verduyn. 1993. *Introduction to Functional Differential Equations*. Vol. 99. Springer Science & Business Media.

Hongfei, L. 2015 (July). Refined stability of a class of CDFE with distributed delays. *Pages 1435–1440 of: 2015 34th Chinese Control Conference (CCC)*.

Kharitonov, Vladimir L, Mondié, Sabine, & Ochoa, Gilberto. 2009. Frequency stability analysis of linear systems with general distributed delays. *Pages 25–36 of: Topics in Time Delay Systems*. Springer.

Li, Hongfei. 2012. Discretized LKF method for stability of coupled differential-difference equations with multiple discrete and distributed delays. *International Journal of Robust and Nonlinear Control*, **22**(8), 875–891.

Löfberg, Johan. 2004. YALMIP: A toolbox for modeling and optimization in MATLAB. *Pages 284–289 of: Computer Aided Control Systems Design, 2004 IEEE International Symposium on*. IEEE.

Michiels, Wim, & Niculescu, Silviu-Iulian. 2014. *Stability, Control, and Computation for Time-Delay Systems: An Eigenvalue-Based Approach*. Vol. 27. SIAM.

Mosek, ApS. 2016. MOSEK MATLAB Toolbox. *Release 8.0.0.42*.

Scherer, Carsten, Gahinet, Pascal, & Chilali, Mahmoud. 1997. Multiobjective output-feedback control via LMI optimization. *Automatic Control, IEEE Transactions on*, **42**(7), 896–911.

Scherer, Carsten W, & Hol, Camile WJ. 2006. Matrix sum-of-squares relaxations for robust semi-definite programs. *Mathematical programming*, **107**(1-2), 189–211.

Seuret, Alexandre, & Gouaisbaut, Frédéric. 2014. Complete Quadratic Lyapunov functionals using Bessel-Legendre inequality. *Pages 448–453 of: Control Conference (ECC), 2014 European*. IEEE.

Seuret, Alexandre, & Gouaisbaut, Frédéric. 2015. Hierarchy of LMI conditions for the stability analysis of time-delay systems. *Systems & Control Letters*, **81**, 1–7.

Seuret, Alexandre, Gouaisbaut, Frédéric, & Ariba, Yassine. 2015. Complete quadratic Lyapunov functionals for distributed delay systems. *Automatica*, **62**, 168–176.

Vyhlídal, Tomáš, & Zítek, Pavel. 2014. *QPmR - Quasi-Polynomial Root-Finder: Algorithm Update and Examples*. Cham: Springer International Publishing. Pages 299–312.

Zhang, X.-M., Han Q.-L. 2009. A new stability criterion for a partial element equivalent circuit model of neutral type. *IEEE Transactions on Circuits and Systems II: Express Briefs*, **56**(10), 798–802. cited By 14.